\documentstyle[12pt]{article}
\input epsf

\begin{document}

\begin{titlepage}

\begin{flushright}
UCLA-00-TEP-20\\
UM-TH-00-09\\
June 2000
\end{flushright}
\vspace{2.0cm}

\begin{center}
\large\bf
{\LARGE\bf Reduction and evaluation of two-loop graphs 
           with arbitrary masses}\\[2cm]
\rm
{Adrian Ghinculov$^a$ and York-Peng Yao$^b$}\\[.5cm]

{\em $^a$Department of Physics and Astronomy, UCLA,}\\
      {\em Los Angeles, California 90095-1547, USA}\\[.1cm]
{\em $^b$Randall Laboratory of Physics, University of Michigan,}\\
      {\em Ann Arbor, Michigan 48109-1120, USA}\\[3.0cm]
      
\end{center}
\normalsize

\begin{abstract}
We describe a general analytic-numerical reduction scheme for evaluating 
any two-loop diagrams with general kinematics and general 
renormalizable interactions, whereby ten special functions form a
complete set after tensor reduction. We discuss the symmetrical 
analytic structure of these special functions in their integral 
representation, which allows for optimized numerical integration.  
The process $Z \to b \bar b $ is used for illustration, 
for which we evaluate all the three-point, non-factorizable $g^2 \alpha_s$
mixed electroweak-QCD graphs, which depend on the top 
quark mass. The isolation of infrared singularities
is detailed, and numerical results are given for all two-loop three-point 
graphs  involved in this process.
\end{abstract}

\vspace{3cm}

\end{titlepage}


\title{Reduction and evaluation of two-loop graphs 
           with arbitrary masses}

\author{Adrian Ghinculov$^a$ and York-Peng Yao$^b$}

\date{{\em $^a$Department of Physics and Astronomy, UCLA,}\\
      {\em Los Angeles, California 90095-1547, USA}\\
      {\em $^b$Randall Laboratory of Physics, University of Michigan,}\\
      {\em Ann Arbor, Michigan 48109-1120, USA}}

\maketitle

\begin{abstract}
We describe a general analytic-numerical reduction scheme for evaluating 
any two-loop diagrams with general kinematics and general 
renormalizable interactions, whereby ten special functions form a
complete set after tensor reduction. We discuss the symmetrical 
analytic structure of these special functions in their integral 
representation, which allows for optimized numerical integration.  
The process $Z \to b \bar b $ is used for illustration, 
for which we evaluate all the three-point, non-factorizable $g^2 \alpha_s$
mixed electroweak-QCD graphs, which depend on the top 
quark mass. The isolation of infrared singularities
is detailed, and numerical results are given for all two-loop three-point
graphs  involved in this process.
\end{abstract}


\section{Introduction}

The success of the perturbative aspect of the Standard Model 
is truly impressive, the theory being now tested at higher 
than one-loop accuracy.  It is natural that over the 
past years various methods have been proposed to calculate 
higher loop graphs.
Beyond one loop, there are  
situations in which one can neglect internal masses, or
exploit the kinematics to evaluate 
amplitudes analytically at some special point where the Feynman diagrams 
become simpler.  
These special situations, while ingenious, cover only certain 
classes of physical processes.  

There have been attempts to formulate solutions for two-loop Feynman 
diagrams, which ideally should be as general as the one-loop solutions are.
For the massless case a lot of progress has been reached recently \cite{qcd}.
On the other hand, for the general massive case, it has become clear
that some numerical treatment appears unavoidable because of the 
complexity of the scalar integrals involved.

In refs. \cite{weiglein:1,weiglein:2} a reduction scheme was found for treating
massive self-energy diagrams at two-loop, with the resulting master scalar
integrals being evaluated numerically.

In ref. \cite{2loopgeneral} we investigated the problem of a general
massive two-loop algorithm, which would deal with multi-leg diagrams as well.
Once we subscribe to a semi-numerical approach, some demands 
must be made, which are based on the generality, the universality, 
the effectiveness and the accuracy that a particular formalism 
engenders.  We have shown that for any external kinematics 
and internal masses, we can reduce every two-loop amplitude for 
a process due to any renormalizable interactions into a set of 
ten scalar integrals and their derivatives.  

It is not hard to 
see how a result like this is obtained if we consider first a scalar 
theory with trivial interactions \cite{2loopnumerical}.  
Then, there would be two internal
momenta, $p$ and $q$, and by the use of Feynman parameters to combine 
sets of denominators in which either $p$ or $q$ or $p+q$ runs through, and 
to make shifts in 
$p$ and $q$, we have an integrand proportional to 

\begin{equation}
1 \over [(p+k)^2+m_1^2]^{\alpha_1}(q^2+m_2^2)^{\alpha_2}
[(p+q)^2+m_3^2]^{\alpha_3}
\end{equation}
in which a four-vector $k$ is left over, which is a linear function 
of the external momenta and Feynman parameters.  The ``masses''
$m_{1,2,3}^2$ are functions of external momenta, internal 
masses and Feynman parameters.  These are to be integrated over 
$p$, $q$ and a set of Feynman parameters.  
By partial differentiating with respect to $m_{1,2,3}^2$, 
there is in principle only one basic function arising 
from $\alpha_1=\alpha_2=\alpha_2=1$ we need to know, although for convenience 
one may add $\alpha_1=2, \alpha_2=\alpha_3=1.$   It is clear 
that different interactions and different graphs will give 
different polynomials of Feynman parameters to the numerators, and 
also different $k$ and $m_{1,2,3}^2$.  An extension of this 
construction to tensor integrals
will give us the main result mentioned earlier.

In the following section, we discuss  the steps and 
arguments needed to perform a tensorial decomposition.  
We  show that there is a small set of basic functions which are needed, 
which have simple, one-dimensional integral representations.
Their analytic structure 
is easy to see, and therefore the integration contour can be 
extended into the complex plane and optimized for 
rapid numerical convergence.  In Section 3, we shall give a detailed 
exposition of the numerical techniques involved. 

Given that two-point functions are usually simpler to calculate,
many calculations existing in the literature used unitarity cuts of
self-energies to calculate lower loop-order inclusive decay rates.
We would like to stress that, given a process, our formalism allows 
us to calculate the amplitudes individually, as opposed to using unitarity 
cuts.  As an immediate 
consequence, it becomes possible  to obtain rates for exclusive 
processes, without the need to integrate over the whole final-state 
phase space. 

In section 4, for illustration, 
we discuss in detail the application of our method to
the important physical process  $Z \to b \bar  b$.

\section{Analytical reduction}

\subsection{Relation with sunset-type integrals}

\begin{figure}
    \epsfxsize = 12cm
\begin{center}
    \epsffile{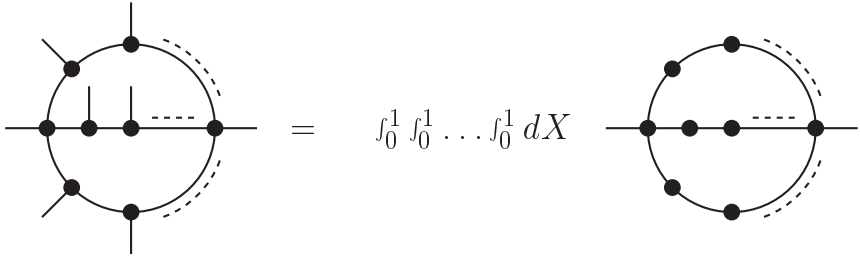}
\end{center}
\caption{Expressing generic massive two-loop Feynman diagrams as integrals
         over sunset-type functions.}
\label{fig:intosunset}
\end{figure}

The starting point of calculating a massive two-loop diagram is to express
it in terms of integrals of a standard type. 
Topologically, any generic two-loop
diagram can be represented in the form shown in figure 1. By introducing
a set of Feynman parameters $X$ to combine
all propagators which have the same loop integration momentum $p$, $q$, or
$r=p+q$, the graph can always be represented as an integral over a sunset-type
two-loop integral. This is illustrated in figure~\ref{fig:intosunset}. 
All dependence on the
external momenta $k_i$ and internal masses $m_i$ is now contained in the
variables $m_1^2$, $m_2^2$, $m_3^2$, and $k$, which also depend on the set
of Feynman parameters $X$.

The original Feynman diagram is written at this point as an integral over
tensor functions of the following type (all momenta are rotated 
into Euclidean):

\begin{equation}
   \int d^{n}p\,d^{n}q\, 
       \frac{p^{\mu_1} \ldots p^{\mu_i} q^{\mu_{i+1}} \ldots q^{\mu_j}}{
             [(p+k)^{2}+m_{1}^{2}]^{\alpha_{1}} \,
             (q^{2}+m_{2}^{2})^{\alpha_{2}} \,
             (r^{2}+m_{3}^{2})^{\alpha_{3}}
	    }
    ~~ .
\label{eq:1}
\end{equation}

By casting the graph in this form, our strategy is to develop a uniform
treatment of all possible sunset-type functions which can be generated
from generic two-loop Feynman graphs. The further tensor reduction and 
decomposition into standard scalar functions is done by using standard
formulae common for all diagrams. Therefore it can be automatized in the
form of an algebraic manipulation program. 

At the end, the remaining 
integral over the Feynman parameters $X$, represented in 
figure~\ref{fig:intosunset}, is performed numerically.

\subsection{Tensor reduction}

Tensor integrals of the type in eqn.~\ref{eq:1} need to be decomposed
into scalar integrals. By Lorentz covariance, this two-loop integral 
is a tensor constructed from the external momentum $k^{\mu}$ and the
metric tensor $g^{\mu\nu}$.
Given that, one way of obtaining the tensor decomposition would be to
write down all tensor structures allowed by the symmetry of the integral, 
use appropriate projectors, and solve the resulting equations. 

Another way to do this is by decomposing
the loop momenta $p$ and $q$ into components parallel and orthogonal to the
external momentum $k^{\mu}$: 

\begin{equation}
 p_{\perp}^{\mu}  =   p^{\mu} - \frac{p \cdot k}{k^2} k^{\mu} 
  ~~,~~
 q_{\perp}^{\mu}  =   q^{\mu} - \frac{q \cdot k}{k^2} k^{\mu} 
    ~~ .
\end{equation}
After this decomposition, the tensor decomposition is obtained by
noticing that the functions with an odd number of transverse loop momenta
$p_{\perp}^{\mu}$ and $q_{\perp}^{\mu}$ vanish, while the even functions
are transverse to $k^{\mu}$ \cite{2loopgeneral}.
In ref. \cite{2loopgeneral} 
we have shown that the resulting
scalar coefficients of the tensor decomposition are integrals of the
following form:

\begin{equation}
\tilde {\cal P}^{a \, b}_{\alpha_1 \, \alpha_2 \, \alpha_3} (m_1,m_2,m_3;k^2)
  =
    \int d^{n}p\,d^{n}q\, 
       \frac{(p \cdot k)^a (q \cdot k)^b}{
             [(p+k)^{2}+m_{1}^{2}]^{\alpha_{1}} \,
             (q^{2}+m_{2}^{2})^{\alpha_{2}} \,
             (r^{2}+m_{3}^{2})^{\alpha_{3}}
	    }
    ~~ .
\label{eq:2}
\end{equation}

We give in the following the tensor decompositions for all 
tensor integrals of the type of eq.~\ref{eq:1} of rank 1, 2, and 3:

\begin{eqnarray}
\frac{1}{[211]} & = &               {_1A_1}  
 ~~~~~~~~~~~~~~~~~~,~
\frac{p^{\mu}}{[211]}  =  k^{\mu} {_2A_1}   
 ~~~~~~~~~~~~~~,~
\frac{q^{\mu}}{[211]}  =  k^{\mu} {_3A_1}   
  \nonumber \\
\frac{p^{\mu} p^{\nu}}{[211]}  & = & \tau^{\mu\nu} {_4A_1} + g^{\mu\nu} {_4A_2}  
 ~,~
\frac{p^{\mu} q^{\nu}}{[211]}   =  \tau^{\mu\nu} {_5A_1} + g^{\mu\nu} {_5A_2}  
 ~,~
\frac{q^{\mu} q^{\nu}}{[211]}   =  \tau^{\mu\nu} {_6A_1} + g^{\mu\nu} {_6A_2}  
  \nonumber \\
\frac{p^{\mu} p^{\nu} p^{\lambda}}{[211]}  & = & 
               ( \tau^{\mu\nu} k^{\lambda} + \tau^{\mu\lambda} k^{\nu} 
                                     + \tau^{\nu\lambda} k^{\mu} ) {_7A_1}  
              + ( g^{\mu\nu} k^{\lambda} + g^{\mu\lambda} k^{\nu} 
                                     + g^{\nu\lambda} k^{\mu} )    {_7A_2} 
  \nonumber \\
\frac{q^{\mu} p^{\nu} p^{\lambda}}{[211]}  & = & 
               ( \tau^{\mu\nu} k^{\lambda} + \tau^{\mu\lambda} k^{\nu} 
                                     + \tau^{\nu\lambda} k^{\mu} ) {_8A_1} 
              + ( g^{\mu\nu} k^{\lambda} + g^{\mu\lambda} k^{\nu} 
                                     + g^{\nu\lambda} k^{\mu} )    {_8A_2}
  \nonumber \\
\frac{p^{\mu} q^{\nu} q^{\lambda}}{[211]}  & = & 
               ( \tau^{\mu\nu} k^{\lambda} + \tau^{\mu\lambda} k^{\nu} 
                                     + \tau^{\nu\lambda} k^{\mu} ) {_9A_1} 
              + ( g^{\mu\nu} k^{\lambda} + g^{\mu\lambda} k^{\nu} 
                                     + g^{\nu\lambda} k^{\mu} )    {_9A_2} 
  \nonumber \\
  & &       
 \;\;\;\;\;\;\;\;\;\;\;\;\;\;\;\;\;\;\;\;\;\;\;\;\;
 \;\;\;\;\;\;\;\;\;\;\;\;\;\;\;\;\;\;
             + ( g^{\mu\nu} k^{\lambda} + g^{\mu\lambda} k^{\nu} 
                                  - 2 g^{\nu\lambda} k^{\mu} )    {_9A_3} 
  \nonumber \\
\frac{q^{\mu} q^{\nu} q^{\lambda}}{[211]}  & = & 
               ( \tau^{\mu\nu} k^{\lambda} + \tau^{\mu\lambda} k^{\nu} 
                                     + \tau^{\nu\lambda} k^{\mu} ) {_{10}A_1} 
              + ( g^{\mu\nu} k^{\lambda}  + g^{\mu\lambda} k^{\nu}  
                                     + g^{\nu\lambda} k^{\mu} )    {_{10}A_2} 
  \nonumber 
\end{eqnarray}

Tensor decompositions for more than three Lorentz indices are derived in a similar way.
In the formulae above, a loop integration $\int d^n p \; d^n q$ 
is understood, and we used the following notations:

\begin{equation}
[211] = [(p+k)^2+m_1^2]^2 \, (q^2+m_2^2) \, (r^2+m_3^2)
 ~~~,~~~
  \tau^{\mu\nu} = g^{\mu\nu} - \frac{k^{\mu}k^{\nu}}{k^2}
\end{equation}
and:

\begin{eqnarray}
 _1A_1    & = & {\cal H}_1  
 ~~~~~~~~~~~~~~~~~~~~~~~~~,~~~~~~~
 _2A_1     =   \frac{1}{k^2}  {\cal H}_2  
 ~~~~~~~~~~~~~,~~~
 _3A_1     =   \frac{1}{k^2}  {\cal H}_3  
  \nonumber \\
 _4A_1    & = &  - \frac{1}{k^2} \frac{n}{n-1}  {\cal H}_4       
 ~~~~~~~~~~~~,~~~~~~
 _4A_2     =     \frac{1}{k^2} \tilde{{\cal P}}^{20}_{211} 
  \nonumber \\
 _5A_1    & = &  - \frac{1}{k^2} \frac{n}{n-1}  {\cal H}_5       
 ~~~~~~~~~~~~,~~~~~~
 _5A_2     =     \frac{1}{k^2}   \tilde{{\cal P}}^{11}_{211} 
  \nonumber \\
 _6A_1    & = &  - \frac{1}{k^2} \frac{n}{n-1}  {\cal H}_6       
 ~~~~~~~~~~~~,~~~~~~
 _6A_2     =     \frac{1}{k^2}   \tilde{{\cal P}}^{02}_{211} 
  \nonumber \\
 _7A_1    & = & -  \left(\frac{1}{k^2}\right)^2 \frac{n+2}{3(n-1)} {\cal H}_7       
 ~~,~~~~~~
 _7A_2     =     \left(\frac{1}{k^2}\right)^2 \frac{1}{3}  
                   \tilde{{\cal P}}^{30}_{211} 
  \nonumber \\
 _8A_1    & = & -  \left(\frac{1}{k^2}\right)^2 \frac{n+2}{3(n-1)} {\cal H}_8       
 ~~,~~~~~~
 _8A_2     =     \left(\frac{1}{k^2}\right)^2 \frac{1}{3}  
                   \tilde{{\cal P}}^{21}_{211} 
  \nonumber \\
 _9A_1    & = & -  \left(\frac{1}{k^2}\right)^2  \frac{n+2}{3(n-1)} {\cal H}_9       
 ~~,~~~~~~
 _9A_2     =     \left(\frac{1}{k^2}\right)^2 \frac{1}{3}  
                   \tilde{{\cal P}}^{12}_{211} 
  \nonumber \\
 _9A_3    & = &    \frac{1}{k^2} \frac{1}{3}    
              \left[ \tilde{{\cal P}}^{11}_{211} + 
                      \tilde{{\cal P}}^{02}_{211} - \frac{n}{n-1} ({\cal H}_5 + {\cal H}_6) 
             \right] 
  \nonumber \\
 _{10}A_1 & = & -  \left(\frac{1}{k^2}\right)^2 \frac{n+2}{3(n-1)} {\cal H}_{10}      
 ~~,~~~~
 _{10}A_2  =     \left(\frac{1}{k^2}\right)^2 \frac{1}{3}  
                   \tilde{{\cal P}}^{03}_{211} 
  \nonumber 
\end{eqnarray}

In the expressions above, we introduced a set of ten scalar integrals
${\cal H}_i$, which are related to the $\tilde{{\cal P}}^{ab}_{211}$
functions, and are defined as follows:

\begin{eqnarray}
    {\cal H}_1 
   & = &
    \int d^{n}p\,d^{n}q\, 
    \frac{1}{
          [(p+k)^2+m_{1}^{2}]^2 \,
          (q^2    +m_{2}^{2}) \,
          (r^2    +m_{3}^{2})         }   
  \nonumber \\
    {\cal H}_2
   & = &
    \int d^{n}p\,d^{n}q\, 
    \frac{ p \cdot k }{
          [(p+k)^2+m_{1}^{2}]^2 \,
          (q^2    +m_{2}^{2}) \,
          (r^2    +m_{3}^{2})         }   
  \nonumber \\
    {\cal H}_3
   & = &
    \int d^{n}p\,d^{n}q\, 
    \frac{ q \cdot k }{
          [(p+k)^2+m_{1}^{2}]^2 \,
          (q^2    +m_{2}^{2}) \,
          (r^2    +m_{3}^{2})         }   
  \nonumber \\
    {\cal H}_4
   & = &
    \int d^{n}p\,d^{n}q\, 
    \frac{ (p \cdot k)^2 
         - \frac{1}{n} k^2 p^2 }{
          [(p+k)^2+m_{1}^{2}]^2 \,
          (q^2    +m_{2}^{2}) \,
          (r^2    +m_{3}^{2})         }   
  \nonumber \\
    {\cal H}_5
   & = &
    \int d^{n}p\,d^{n}q\, 
    \frac{ (p \cdot k) (q \cdot k) 
         - \frac{1}{n} k^2 (q \cdot p)  }{
          [(p+k)^2+m_{1}^{2}]^2 \,
          (q^2    +m_{2}^{2}) \,
          (r^2    +m_{3}^{2})         }   
  \nonumber \\
    {\cal H}_6
   & = &
    \int d^{n}p\,d^{n}q\, 
    \frac{ (q \cdot k)^2 
         - \frac{1}{n} k^2 q^2  }{
          [(p+k)^2+m_{1}^{2}]^2 \,
          (q^2    +m_{2}^{2}) \,
          (r^2    +m_{3}^{2})         }   
  \nonumber \\
    {\cal H}_7
   & = &
    \int d^{n}p\,d^{n}q\, 
    \frac{ (p \cdot k)^3 
         - \frac{3}{n+2} k^2 p^2 (p \cdot k)  }{
          [(p+k)^2+m_{1}^{2}]^2 \,
          (q^2    +m_{2}^{2}) \,
          (r^2    +m_{3}^{2})         }   
  \nonumber \\
   {\cal H}_8
   & = &
    \int d^{n}p\,d^{n}q\, 
    \frac{ (p \cdot k)^2 (q \cdot k) 
         -   \frac{3}{n+2} k^2 p^2 (q \cdot k)   }{
          [(p+k)^2+m_{1}^{2}]^2 \,
          (q^2    +m_{2}^{2}) \,
          (r^2    +m_{3}^{2})         }   
  \nonumber \\
    {\cal H}_9 
   & = &
    \int d^{n}p\,d^{n}q\, 
    \frac{ (p \cdot k) (q \cdot k)^2 
        -  \frac{1}{n+2} k^2 
	   [ 2 (p \cdot q) (q \cdot k) + q^2 (p \cdot k) ]   }{
          [(p+k)^2+m_{1}^{2}]^2 \,
          (q^2    +m_{2}^{2}) \,
          (r^2    +m_{3}^{2})         }   
  \nonumber \\
    {\cal H}_{10} 
   & = &
    \int d^{n}p\,d^{n}q\, 
    \frac{ (q \cdot k)^3 
         - \frac{3}{n+2}  k^2 q^2 (q \cdot k)  }{
          [(p+k)^2+m_{1}^{2}]^2 \,
          (q^2    +m_{2}^{2}) \,
          (r^2    +m_{3}^{2})         }   
\end{eqnarray}

As it can be seen in eqs. 5, these scalar functions appear naturally 
in the tensorial decomposition. As discussed in the following section, the
functions ${\cal H}_i$ are only logarithmically divergent in the ultraviolet.
Because of this, they have very simple integral representations which can be 
used for numerical computation. Once a way of calculating the special
functions ${\cal H}_i$ is available, the $\tilde {\cal P}^{ab}_{2 \, 1 \, 1}$
functions can easily be recovered by partial fractioning eqns. 6. 
The partial fractioning generates essentially trivial products of one-loop
tadpoles. The conversion formulae are given in Appendix A.

\subsection{Integral representation of the ${\cal H}_i$ functions}

Because the ultraviolet behaviour of the functions ${\cal H}_i$ is 
logarithmic, fairly simple and symmetric integral representations can be 
found \cite{2loopgeneral}:

\begin{eqnarray}
   {\cal H}_1 
   & = &
    \pi^4 
    \left[
      \frac{2}{\epsilon^2}
    - \frac{1}{\epsilon} ( 1 - 2 \gamma_{m_1} )
          - \frac{1}{2}
          + \frac{\pi^2}{12}
          - \gamma_{m_1}
          + \gamma_{m_1}^2
          + h_1
   \right]
  \nonumber \\
   {\cal H}_2 
   & = &
    \pi^4 k^2
    \left[
    - \frac{2}{\epsilon^2}
    + \frac{1}{\epsilon} ( \frac{1}{2} - 2 \gamma_{m_1} )
          + \frac{13}{8}
          - \frac{\pi^2}{12}
          + \frac{\gamma_{m_1}}{2}
          - \gamma_{m_1}^2
          - h_2
   \right]
  \nonumber \\
   {\cal H}_3 
   & = &
    \pi^4 k^2
    \left[
      \frac{1}{\epsilon^2}
    - \frac{1}{\epsilon} ( \frac{1}{4} - \gamma_{m_1} )
          - \frac{13}{16}
          + \frac{\pi^2}{24}
          - \frac{\gamma_{m_1}}{4} 
          + \frac{\gamma_{m_1}^2}{2} 
          + h_3
   \right]
  \nonumber \\
   {\cal H}_4 
   & = &
    \pi^4 (k^2)^2
    \left[
      \frac{3}{2 \epsilon^2}
    + \frac{1}{\epsilon} \frac{3 \gamma_{m_1}}{2} 
          - \frac{175}{96}
          + \frac{\pi^2}{16}
          + \frac{3 \gamma_{m_1}^2}{4} 
          + \frac{3}{4} h_4
   \right]
  \nonumber \\
   {\cal H}_5 
   & = &
    \pi^4 (k^2)^2
    \left[
    - \frac{3}{4 \epsilon^2}
    - \frac{1}{\epsilon} \frac{3 \gamma_{m_1}}{4} 
          + \frac{175}{192}
          - \frac{\pi^2}{32}
          - \frac{3 \gamma_{m_1}^2}{8} 
          - \frac{3}{4} h_5
   \right]
  \nonumber \\
   {\cal H}_6 
   & = &
    \pi^4 (k^2)^2
    \left[
      \frac{1}{2 \epsilon^2}
    - \frac{1}{\epsilon} ( \frac{1}{24} - \frac{\gamma_{m_1}}{2}  )
          - \frac{19}{32}
          + \frac{\pi^2}{48}
          - \frac{\gamma_{m_1}}{24} 
          + \frac{\gamma_{m_1}^2}{4} 
          + \frac{3}{4} h_6
   \right]
  \nonumber \\
   {\cal H}_7
   & = &
    \pi^4 (k^2)^3
    \left[
    - \frac{1}{\epsilon^2}
    - \frac{1}{\epsilon} ( \frac{5}{24} + \gamma_{m_1} )
          + \frac{287}{192}
          - \frac{\pi^2}{24}
          - \frac{5 \gamma_{m_1}}{24} 
          - \frac{\gamma_{m_1}^2}{2} 
          - \frac{1}{2} h_7
   \right]
  \nonumber \\
   {\cal H}_8
   & = &
    \pi^4 (k^2)^3
    \left[
      \frac{1}{2\epsilon^2}
    + \frac{1}{\epsilon} ( \frac{5}{48} + \frac{\gamma_{m_1}}{2}  )
          - \frac{287}{384}
          + \frac{\pi^2}{48}
          + \frac{5 \gamma_{m_1}}{48} 
          + \frac{\gamma_{m_1}^2}{4} 
          + \frac{1}{2} h_8
   \right]
  \nonumber \\
   {\cal H}_9 
   & = &
    \pi^4 (k^2)^3
    \left[
    - \frac{1}{3 \epsilon^2}
    - \frac{1}{\epsilon} ( \frac{1}{24} + \frac{\gamma_{m_1}}{3}  )
          + \frac{95}{192}
          - \frac{\pi^2}{72}
          - \frac{\gamma_{m_1}}{24} 
          - \frac{\gamma_{m_1}^2}{6} 
          - \frac{1}{2} h_9
   \right]
  \nonumber \\
   {\cal H}_{10} 
   & = &
    \pi^4 (k^2)^3
    \left[
      \frac{1}{4 \epsilon^2}
    + \frac{1}{\epsilon} ( \frac{1}{96} + \frac{\gamma_{m_1}}{4}  )
          - \frac{283}{768}
          + \frac{\pi^2}{96}
          + \frac{\gamma_{m_1}}{96} 
          + \frac{\gamma_{m_1}^2}{8} 
          + \frac{1}{2} h_{10}
   \right]  
   \; \; .
\end{eqnarray}

Here, $n=4+\epsilon$ is the space-time dimension, and 
$\gamma_m=\gamma + \log(\pi m^2/\mu_1^2)$ ($\gamma$ is the Euler constant, 
and $\mu_1$ is the 't Hooft mass).
The special functions $h_i$ which appear in the formulae above are 
the finite part in the $1/\epsilon$ expansion of ${\cal H}_i$. 
They are defined by the following integral representations. 
Except for special values of their arguments, they cannot be 
further integrated into well-studied functions, such as the 
familiar polylogarithms, and as such, our strategy is to evaluate 
them numerically directly from their integral representations:

\begin{eqnarray}
    h_1(m_1,m_2,m_3;k^2) & = &  \int_0^1 dx \,
                                \tilde{g} (x)
  \nonumber \\
    h_2(m_1,m_2,m_3;k^2) & = &  \int_0^1 dx \,
                              [ \tilde{g}   (x)
                              + \tilde{f_1} (x) ]
  \nonumber \\
    h_3(m_1,m_2,m_3;k^2) & = &  \int_0^1 dx \, 
                              [ \tilde{g}   (x)
                              + \tilde{f_1} (x) ] \, (1-x)
  \nonumber \\
    h_4(m_1,m_2,m_3;k^2) & = &  \int_0^1 dx \,
                              [ \tilde{g}   (x)
                              + \tilde{f_1} (x)
                              + \tilde{f_2} (x) ]
  \nonumber \\
    h_5(m_1,m_2,m_3;k^2) & = &  \int_0^1 dx \,
                              [ \tilde{g}   (x)
                              + \tilde{f_1} (x)
                              + \tilde{f_2} (x) ] \, (1-x)
  \nonumber \\
    h_6(m_1,m_2,m_3;k^2) & = &  \int_0^1 dx \,
                              [ \tilde{g}   (x)
                              + \tilde{f_1} (x)
                              + \tilde{f_2} (x) ] \, (1-x)^2
  \nonumber \\
    h_7(m_1,m_2,m_3;k^2) & = &  \int_0^1 dx \,
                              [ \tilde{g}   (x)
                              + \tilde{f_1} (x)
                              + \tilde{f_2} (x)
                              + \tilde{f_3} (x) ]
  \nonumber \\
    h_8(m_1,m_2,m_3;k^2) & = &  \int_0^1 dx \,
                              [ \tilde{g}   (x)
                              + \tilde{f_1} (x)
                              + \tilde{f_2} (x)
                              + \tilde{f_3} (x) ] \, (1-x)
  \nonumber \\
    h_9(m_1,m_2,m_3;k^2) & = &  \int_0^1 dx \,
                              [ \tilde{g}   (x)
                              + \tilde{f_1} (x)
                              + \tilde{f_2} (x)
                              + \tilde{f_3} (x) ] \, (1-x)^2
  \nonumber \\
    h_{10}(m_1,m_2,m_3;k^2) & = &  \int_0^1 dx \,
                              [ \tilde{g}   (x)
                              + \tilde{f_1} (x)
                              + \tilde{f_2} (x)
                              + \tilde{f_3} (x) ] \, (1-x)^3
   \; \; .
\label{eq:intrepr}
\end{eqnarray}

The four building blocks $\tilde{g}(x)$, $\tilde{f_1}(x)$, $\tilde{f_2}(x)$,
and $\tilde{f_3}(x)$  of these one-dimensional integral representations
are given in Appendix B.

\subsection{Completeness of the $\{{\cal H}_i\}_{i=\overline{1,10}}$ 
            special functions for renormalizable theories}

The ten special functions ${\cal H}_i$ we introduced in the previous
section, or, equivalently, their corresponding 
$\tilde {\cal P}^{a b}_{\alpha_1\alpha_2\alpha_3}$, 
are sufficient for treating all two-loop Feynman graphs which
may be encountered in renormalizable theories. 

To see this, it is useful to define the following auxiliary two-loop functions:

\begin{equation}
{\cal P}^{a \, b}_{\alpha_1 \, \alpha_2 \, \alpha_3} (m_1,m_2,m_3;k^2)
  =
    \int d^{n}p\,d^{n}q\, 
       \frac{(p \cdot k)^a (q \cdot k)^b}{
             (p^{2}+m_{1}^{2})^{\alpha_{1}} \,
             (q^{2}+m_{2}^{2})^{\alpha_{2}} \,
             [(r+k)^{2}+m_{3}^{2}]^{\alpha_{3}}
	    }
    ~~ .
\end{equation}

${\cal P}^{a b}_{\alpha_1\alpha_2\alpha_3}$ are trivially related
to $\tilde {\cal P}^{a b}_{\alpha_1\alpha_2\alpha_3}$
by a simple redefinition of the loop momentum (see Appendix A). 
The only reason for introducing them is for simplifying 
the discussion of this section; 
some recursion relations can be written more compactly in terms 
of these scalar functions.

To prove our assertion that the set of ten $\{{\cal H}_i\}$ functions
is sufficient for the case of renormalizable theories, we notice that
not all functions ${\cal P}^{a b}_{\alpha_1\alpha_2\alpha_3}$ of various indices
are independent. There are recursion relations which relate functions
of different indices. 

We will define the ``degree'' of a 
${\cal P}^{a b}_{\alpha_1\alpha_2\alpha_3}$ function as 
$\alpha_1+\alpha_2+\alpha_3-a-b$. The degree can be increased by
differentiating with respect to the mass variables:

\begin{equation}
{\cal P}^{a \, b}_{\alpha_1+1 \, \alpha_2 \, \alpha_3}(m_1,m_2,m_3;k^2) =
 - \frac{1}{\alpha_1} \frac{\partial}{\partial m_1^2}
{\cal P}^{a \, b}_{\alpha_1   \, \alpha_2 \, \alpha_3}(m_1,m_2,m_3;k^2)
    \; \; ,
\end{equation}
and similarly for $\alpha_2$ and $\alpha_3$. 

Functions of the same degree are related by recursion relations obtained
by differentiating with respect to the external momentum variable $k^{\mu}$:

\begin{eqnarray}
  {\cal P}^{a+1 \, b}_{\alpha_1+1 \, \alpha_2 \, \alpha_3} & = &
  \frac{1}{2 \; \alpha_1} \left[ 2 k^2 \frac{\partial}{\partial k^2} 
                                 - (a+b)
                          \right]
  {\cal P}^{a \, b}_{\alpha_1 \, \alpha_2 \, \alpha_3}
  + \frac{a k^2}{2 \alpha_1} 
  {\cal P}^{a-1 \, b}_{\alpha_1 \, \alpha_2 \, \alpha_3}
\nonumber \\
  {\cal P}^{a \, b+1}_{\alpha_1 \, \alpha_2+1 \, \alpha_3} & = &
  \frac{1}{2 \; \alpha_2} \left[ 2 k^2 \frac{\partial}{\partial k^2} 
                                 - (a+b)
                          \right]
  {\cal P}^{a \, b}_{\alpha_1 \, \alpha_2 \, \alpha_3}
  + \frac{b k^2}{2 \alpha_2} 
  {\cal P}^{a \, b-1}_{\alpha_1 \, \alpha_2 \, \alpha_3}
    \; \; ,
\end{eqnarray}
and

\begin{equation}
   \left[ 2 k^2 \frac{\partial}{\partial k^2} - (a+b)
  \right]
  {\cal P}^{a \, b}_{\alpha_1 \, \alpha_2 \, \alpha_3}  = 
  - 2 \alpha_3
   \left[ k^2
   {\cal P}^{a   \, b  }_{\alpha_1 \, \alpha_2 \, \alpha_3+1} +
   {\cal P}^{a+1 \, b  }_{\alpha_1 \, \alpha_2 \, \alpha_3+1} +
   {\cal P}^{a   \, b+1}_{\alpha_1 \, \alpha_2 \, \alpha_3+1}
  \right]
    \; \; .
\end{equation}

When calculating a specific process, it is thus possible to just focus
on the lowest order ${\cal P}^{a b}_{\alpha_1 \alpha_2 \alpha_3}$
functions involved. All the other can be derived from these by differentiation,
by using the relations above.  

\begin{figure}
    \epsfxsize = 5cm
\begin{center}
    \epsffile{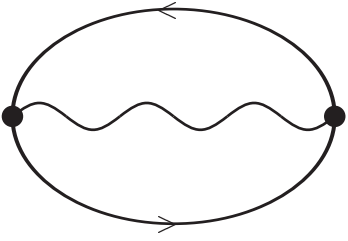}
\end{center}
\caption{Two-loop diagram which involves 
${\cal P}^{a b}_{\alpha_1 \alpha_2 \alpha_3}$ functions of minimal degree one.
It consists of a fermion loop, a boson propagator, 
and non-derivative vertices.}
\label{fig:deg1diagram}
\end{figure}

At the same time, in renormalizable theories, 
there is a lower bound on the possible degree of 
${\cal P}^{a b}_{\alpha_1 \alpha_2 \alpha_3}$ 
functions involved in two-loop graphs, 
imposed by the dimension of the operators
in the Lagrangian. The minimal degree is one, and is attained for instance 
by a two-loop diagram such as the one shown in figure~\ref{fig:deg1diagram}.
Additional external legs can at most increase the degree of the 
${\cal P}^{a b}_{\alpha_1 \alpha_2 \alpha_3}$ functions involved.

Therefore, the most obvious choice of a basic set is the 
${\cal P}^{a b}_{\alpha_1 \alpha_2 \alpha_3}$ functions
with $\alpha_1=\alpha_2=\alpha_3=1$, $a+b=0,1,2$:

\begin{equation}
\begin{array}{llll} 
\mbox{degree} = 3 :                                   \;\;\;\;
                    & {\cal P}^{0 \, 0}_{1 \, 1 \, 1} \;\;\;\;
                    &  
                    &  
\\
\mbox{degree} = 2 :                                   \;\;\;\;
                    & {\cal P}^{1 \, 0}_{1 \, 1 \, 1} \;\;\;\;
                    & {\cal P}^{0 \, 1}_{1 \, 1 \, 1} \;\;\;\;
                    &  
\\
\mbox{degree} = 1 :                                   \;\;\;\;
                    & {\cal P}^{2 \, 0}_{1 \, 1 \, 1} \;\;\;\;
                    & {\cal P}^{1 \, 1}_{1 \, 1 \, 1} \;\;\;\;
                    & {\cal P}^{0 \, 2}_{1 \, 1 \, 1} \;\;\;\;

\end{array}
\label{eq:smallset}
\end{equation}

However, we prefer to use the functions with 
$\alpha_1=2$, $\alpha_2=\alpha_3=1$, $a+b=0,1,2,3$
instead:

\begin{equation}
\begin{array}{lllll} 
\mbox{degree} = 4 :                                   \;\;\;\;
                    & {\cal P}^{0 \, 0}_{2 \, 1 \, 1} \;\;\;\;
                    &  
                    &  
                    &   
\\
\mbox{degree} = 3 :                                   \;\;\;\;
                    & {\cal P}^{1 \, 0}_{2 \, 1 \, 1} \;\;\;\;
                    & {\cal P}^{0 \, 1}_{2 \, 1 \, 1} \;\;\;\;
                    &  
                    &   
\\
\mbox{degree} = 2 :                                   \;\;\;\;
                    & {\cal P}^{2 \, 0}_{2 \, 1 \, 1} \;\;\;\;
                    & {\cal P}^{1 \, 1}_{2 \, 1 \, 1} \;\;\;\;
                    & {\cal P}^{0 \, 2}_{2 \, 1 \, 1} \;\;\;\;
                    &   
\\
\mbox{degree} = 1 :                                   \;\;\;\;
                    & {\cal P}^{3 \, 0}_{2 \, 1 \, 1} \;\;\;\;
                    & {\cal P}^{2 \, 1}_{2 \, 1 \, 1} \;\;\;\;
                    & {\cal P}^{1 \, 2}_{2 \, 1 \, 1} \;\;\;\;
                    & {\cal P}^{0 \, 3}_{2 \, 1 \, 1} \;\;\;\;
\end{array}
\label{eq:largeset}
\end{equation}

It turns out that this equivalent set of functions has simpler
integral representations than the 
functions of eqs.~\ref{eq:smallset}. 
When needed, the eqs.~\ref{eq:smallset}
functions can be derived from  eqs.~\ref{eq:largeset}
by partial $p$:

\begin{eqnarray}
  {\cal P}^{a \, b}_{\alpha_1 \, \alpha_2 \, \alpha_3} 
    & = &
  - \frac{1}{n-(\alpha_1 + \alpha_2 + \alpha_3) + (a+b)/2 }
  \left\{
 \alpha_1 m_1^2  {\cal P}^{a \, b}_{\alpha_1+1 \, \alpha_2 \, \alpha_3} +
 \alpha_2 m_2^2  {\cal P}^{a \, b}_{\alpha_1 \, \alpha_2+1 \, \alpha_3} 
  \right.
\nonumber \\
    &   &
  \left.  \;\;\;\;\;\;\;\;\;\;\;\;\;\;\;\;\;
 + \alpha_3 m_3^2  {\cal P}^{a \, b}_{\alpha_1 \, \alpha_2 \, \alpha_3+1} -
   \frac{1}{2}
   \left[ 2 k^2 \frac{\partial}{\partial k^2} - (a+b)
  \right]
       {\cal P}^{a \, b}_{\alpha_1 \, \alpha_2 \, \alpha_3}
 \right\}
    \; \; ,
\end{eqnarray}

Because the set of functions of eqs.~\ref{eq:largeset} and the ten
${\cal H}_i$ functions differ essentially by simple, one-loop 
tadpole contributions given in Appendix A, this proves our 
assertion that the ten 
${\cal H}_i$ functions are sufficient for treating renormalizable 
theories at two-loop. 

We would like to stress that additional
recursion relations and symmetries exist among this set of functions.
An example of such a symmetry relation is given in 
eq. 16 of ref. \cite{2loopgeneral},
and additional ones may be present. Such relations can be used, in principle,
for restraining the number of master functions involved. However, 
the set of ten ${\cal H}_i$ that we proposed has the advantage of having
clean, symmetrical integral representations, where permutations of mass
arguments are not involved, while symmetry relations obtained by loop
momenta redefinitions often interchange the mass arguments. 
This simplicity is an advantage in practical 
calculations done by computer algebra, where a simpler algorithm may be
preferable to a more complicated one which produces somewhat 
more compact results.

Thus, on general grounds, all tensor integrals ${\cal P}^{ab}_
{\alpha_1 \alpha_2 \alpha_3}$ with $a+b \ge 4$ can be expressed 
by the standard set of ten functions via mass or momentum differentiation.  
The differentiations required can be carried out either by hand, or
automatically, by using a computer algebra program such as Mathematica
or Maple.

However, in practical calculations of Feynman graphs, before performing 
the tensor decomposition, it is advantageous to exploit all possible 
partial fractioning of the loop momenta, instead of resorting to 
recursion relations {\em after} the tensor decomposition. 
This can result in more compact results. This is in particular the case 
with the examples given in section 4. 
There, the calculation can be simplified by noticing that in 
the Feynman gauge, loop momenta $p$ and $q$ in the numerators come from 
fermion propagators, and there are no more than four of them.  
For those terms with four powers of $p$ and $q$, one can 
algebraically rearrange them, so that two of these powers 
will contract as $p^2$, $q^2$ or $p\cdot q$, and then perform 
partial fractioning of these terms, to reduce them into $\tilde P$ 
functions with $a+b\le 3$.

\section{Numerical integration}

Once the tensor reduction is done analytically, the original Feynman
diagram is expressed, according to figure~\ref{fig:intosunset},
as an integral over the set of remaining Feynman parameters ${X}$.
The integrand itself consists of a sum of $h_i$ special functions and
possibly trivial functions such as logarithms and rational functions
of the kinematic variables $m_1^2$, $m_2^2$, $m_3^2$, and $k^2$.

The $h_i$ functions themselves are given by the one-dimensional integral
representations given in eqns.~\ref{eq:intrepr}. 
Within our method, all these integration steps are 
performed numerically in general.

Therefore, it is natural to separate the numerical integration 
into two distinct steps: the routines to calculate the $h_i$ functions
from their integral representations, and the final integration over the
remaining Feynman parameters ${X}$.

When performing the numerical integration, special care is needed to 
make sure that the integration is performed on the physical sheet.
Let us start with the integration of the $h_i$ functions.

\begin{figure}
\begin{center}
    \epsfxsize = 15cm
    \epsffile{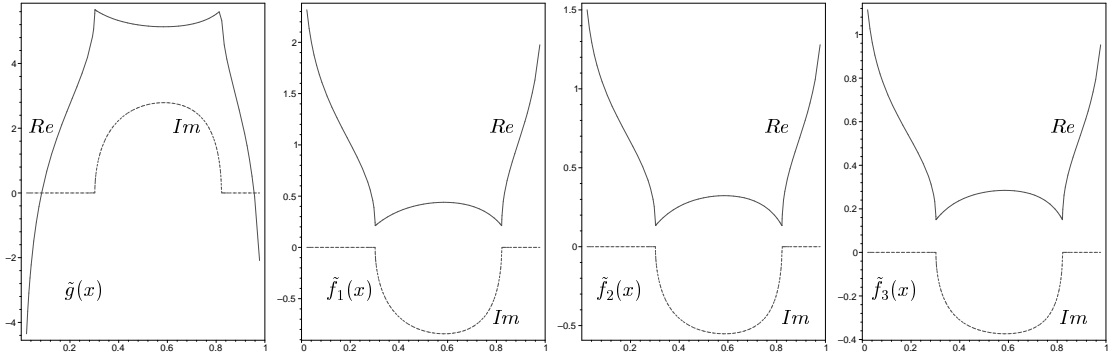}
\end{center}
\caption{Typical behaviour of functions $\tilde g(x)$, $\tilde f_1(x)$,
         $\tilde f_2(x)$, $\tilde f_3(x)$ and of their derivatives
         as a function of the integration parameter $x$. The specific
         kinematic variables in these plots are above threshold, 
         $-k^2 > (m_1+m_2+m3)^2$, where an imaginary part is present.}
\label{fig:gfunction}
\end{figure}

In figure~\ref{fig:gfunction} we plot the functions $\tilde g(x)$ and 
$\tilde f_i(x)$ 
as a function of the integration parameter $x$. 
This is a typical behaviour above the threshold.
At $x=0$ there is an integrable singularity of the logarithmic type, and similarly
at $x=1$. By mass or momentum differentiation the integrable nature of these
singularities in $x$ is not changed; this makes possible the treatment of more complicated
topologies with additional external legs by using the same numerical approach. 
Along the integration path there are two branching points, between
which the function acquires an imaginary part. The integration path must be 
chosen such as to reproduce correctly this imaginary part. 

\begin{figure}[t]
\begin{center}
    \epsfxsize = 14cm
    \epsffile{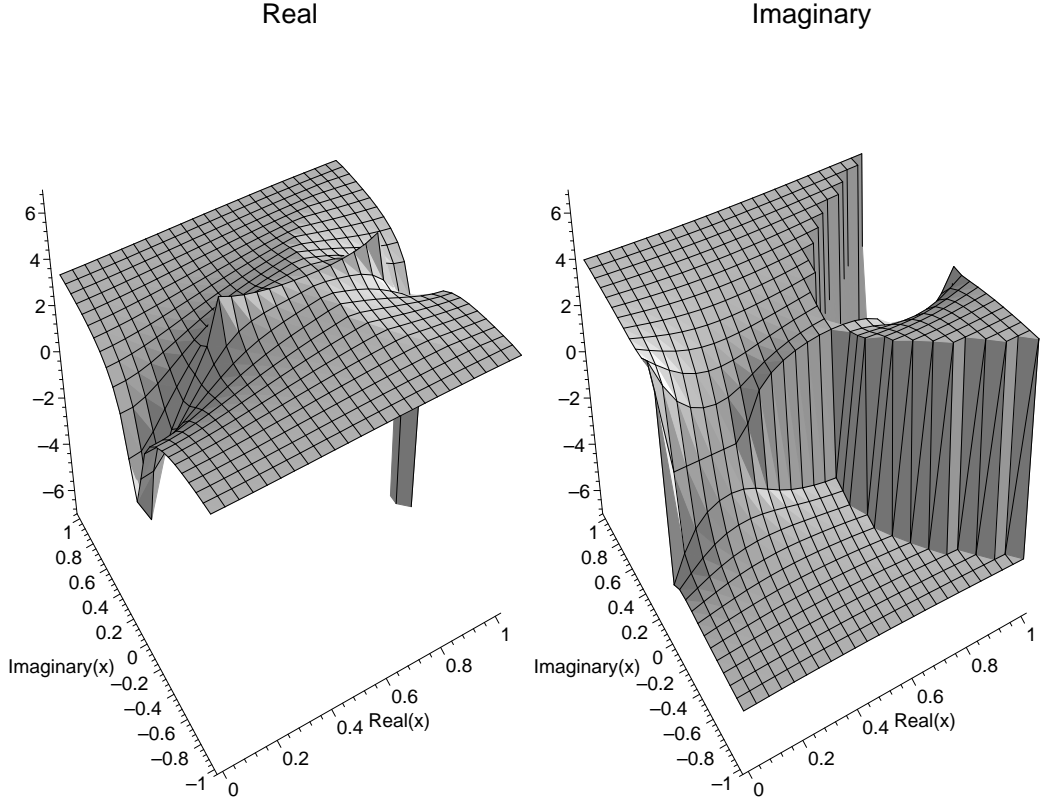}
\end{center}
\caption{Typical behaviour of the functions $\tilde g(x)$, $\tilde f_1(x)$,
         $\tilde f_2(x)$, $\tilde f_3(x)$ and of their derivatives
         in the complex integration parameter $x$. The integration proceeds
         from $x=0$ to $x=1$.}
\label{fig:g3d}
\end{figure}

By continuing the
integration parameter $x$ into the complex plane, we obtain a picture
of the $h_i$ function's integrand as shown in figure~\ref{fig:g3d}.
It is clear that if the position of the two branching points is known,
a correct integration path can be calculated automatically by the computer
program. Finding the singularities involves some subtleties.

Adopting the notations introduced in Appendix B, and inspecting the
expressions of the functions $\tilde g(x)$, $\tilde f_1(x)$, 
$\tilde f_2(x)$ and $\tilde f_3(x)$ given there,
the branching points we are looking for must be among the solutions
of the equation:

\begin{equation}
  \Delta \equiv  (1 + \kappa^{2} - \mu^{2})^{2} 
          + 4 \kappa^{2} \mu^{2} - 4 i \kappa^{2} \eta = 0
     ~~.
\end{equation}

There are four solutions:

\begin{eqnarray}
x_{1,2} & = & \frac{1}{2 \mu_{1}^{2}}
             \, [ \, -a + b + \mu_{1}^{2} 
   \pm \sqrt{(a-b-\mu_{1}^{2})^{2} - 4 b \mu_{1}^{2}} \, ]  \nonumber \\
x_{3,4} & = & \frac{1}{2 \mu_{2}^{2}}
             \, [ \, -a + b + \mu_{2}^{2} 
   \pm \sqrt{(a-b-\mu_{2}^{2})^{2} - 4 b \mu_{2}^{2}} \, ]  
      \; \; ,
                                                         \nonumber \\
\mu_{1,2}^{2}  & = & 1 - \kappa^{2} \mp 2 \sqrt{- \kappa^{2}}
     ~~ .
\label{eq:4solutions}
\end{eqnarray}

To establish which two of the four solutions are the branching
points we are looking for, it is useful to note that the causality
of the Green's functions can be expressed in at least two equivalent
ways, which ought to lead to the same prescription for the integration 
path. The causality condition is expressed by the $i \eta$ prescription 
in the Feynman propagator, which means to shift all masses in the propagators 
$m^{2} \rightarrow m^{2}-i\eta$. An equivalent way to impose causality
is to calculate the Euclidian Green's functions and to go afterwards 
to physical momenta, approaching the cut on the positive real axis 
from above. This amounts to shifting the external momentum
$k^{2} \rightarrow k^{2}+i\eta$. 
These two prescriptions ought to be equivalent, 
and therefore have to fix the location of the physical singularities
with respect to the real axis in the same way.
For $x_{1}$ and $x_{2}$, at $-\kappa^{2} > 1$
both prescriptions lead to the same change, and therefore these
are the singularities of the $\tilde{g}$ and $\tilde f_i$ functions.
For $x_{3}$ and $x_{4}$, the two prescriptions lead to opposite
changes in the imaginary direction. Since causality fixes the location
of the singularities of the Green's function uniquely, 
$x_{3}$ and $x_{4}$ cannot correspond to real singularities of 
$\tilde{g}$ and $\tilde f_i$.
Therefore $\tilde{g}$ and $\tilde f_i$ are analytical at these two points.
The $\tilde{g}$ and $\tilde f_i$ functions themselves have 
only two branching points at $x_{1}$ and $x_{2}$, 
because the singularities at $x_{3}$ and $x_{4}$ are compensating
among the four terms of $\tilde{g}$ in eq. 26. It is interesting to
notice that the spurious solutions $x_{3}$ and $x_{4}$ correspond to
the spurious, unphysical thresholds discussed in ref. \cite{lauricella} in the
context of the relation of the scalar sunset diagram with a generalization
of hypergeometric functions.

\begin{figure}
    \epsfxsize = 10cm
\begin{center}
    \epsffile{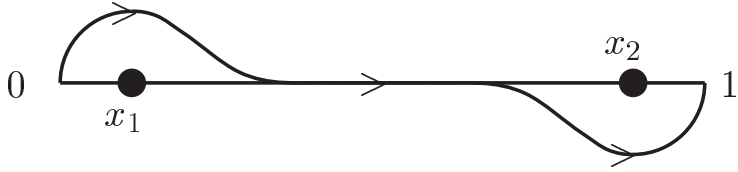}
\end{center}
\caption{A typical shape for the complex integration path, defined in terms
         of higher order spline functions. Both the path and the Jacobian 
         are smooth functions.}
\label{fig:path}
\end{figure}

Once the positions of these two singularities are known, the computer
program can automatically compute an integration path which avoids them.
Because we are interested in a high accuracy and efficiency routine,
we used an adaptive deterministic integration algorithm. Such integration
routines are very accurate provided that the integrand is smooth enough. 
The integrand itself is, of course, an analytic function along the 
complex integration path, and to preserve its smoothness it is advantageous
to define a smooth integration path as well. We use an integration path
defined in terms of higher-order spline functions such that both the
path and its Jacobian are smooth functions. A typical path is shown in 
figure~\ref{fig:path}.

Along these lines, a fully automatic computer program can be written,
which first identifies the singularities, then calculates a suitable
complex integration path, and then performs the numerical integration
to obtain the $h_i$ functions starting from their integral representations
given in eqs.~\ref{eq:intrepr}. By using adaptive integration routines,
one typical evaluation of an $h_i$ function or of a derivative at eight 
digits takes of the order of 30 ms on a 600 MHz Pentium processor.

\begin{figure}
    \epsfxsize = 14cm
\begin{center}
    \epsffile{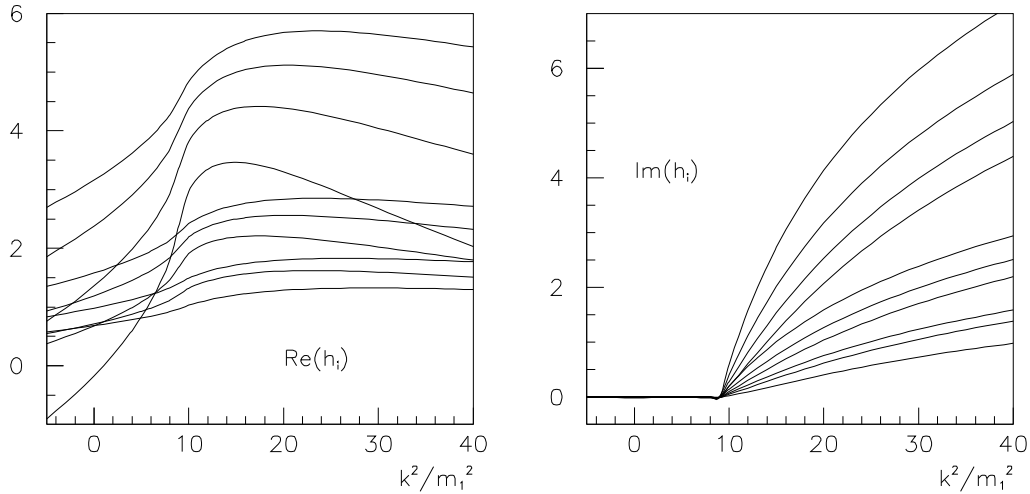}
\end{center}
\caption{Plots of the ten special functions $h_i(m_1^2,m_2^2,m_3^2;-k^2)$ 
         as a function of the external momentum variable $k^2$. The
         plots given here are for $m_1^2=m_2^2=m_3^2=1$.}
\end{figure}

In practical calculations of Feynman graphs, a number of mass or momentum
derivatives of the functions $h_i$ become necessary. They can easily be
obtained either by hand, or automatically, by computer algebra programs such 
as Mathematica or Maple.

The numerical integration over the remaining Feynman parameters ${X}$
is carried out along similar lines \cite{2loopnumerical,onshellexp,ztopdecay}. 
The dimensionality of this final
integration depends on the topology of the diagram and on the number
of legs.

With increasing complexity of the diagram, the methods we describe
in this paper will still be applicable in principle, but in practice will
result in larger integrands and a final numerical integration of higher 
dimension. Thus its applicability will be limited in practice by the available
computing power and by the ability of handling potentially 
large expressions which result from the tensor reduction 
in an error free way.

\section{Examples}

\subsection{Three-point two-loop diagrams contributing to 
            $Z \rightarrow b\bar b$}

\begin{figure}
    \epsfxsize = 14.5cm
\begin{center}
    \epsffile{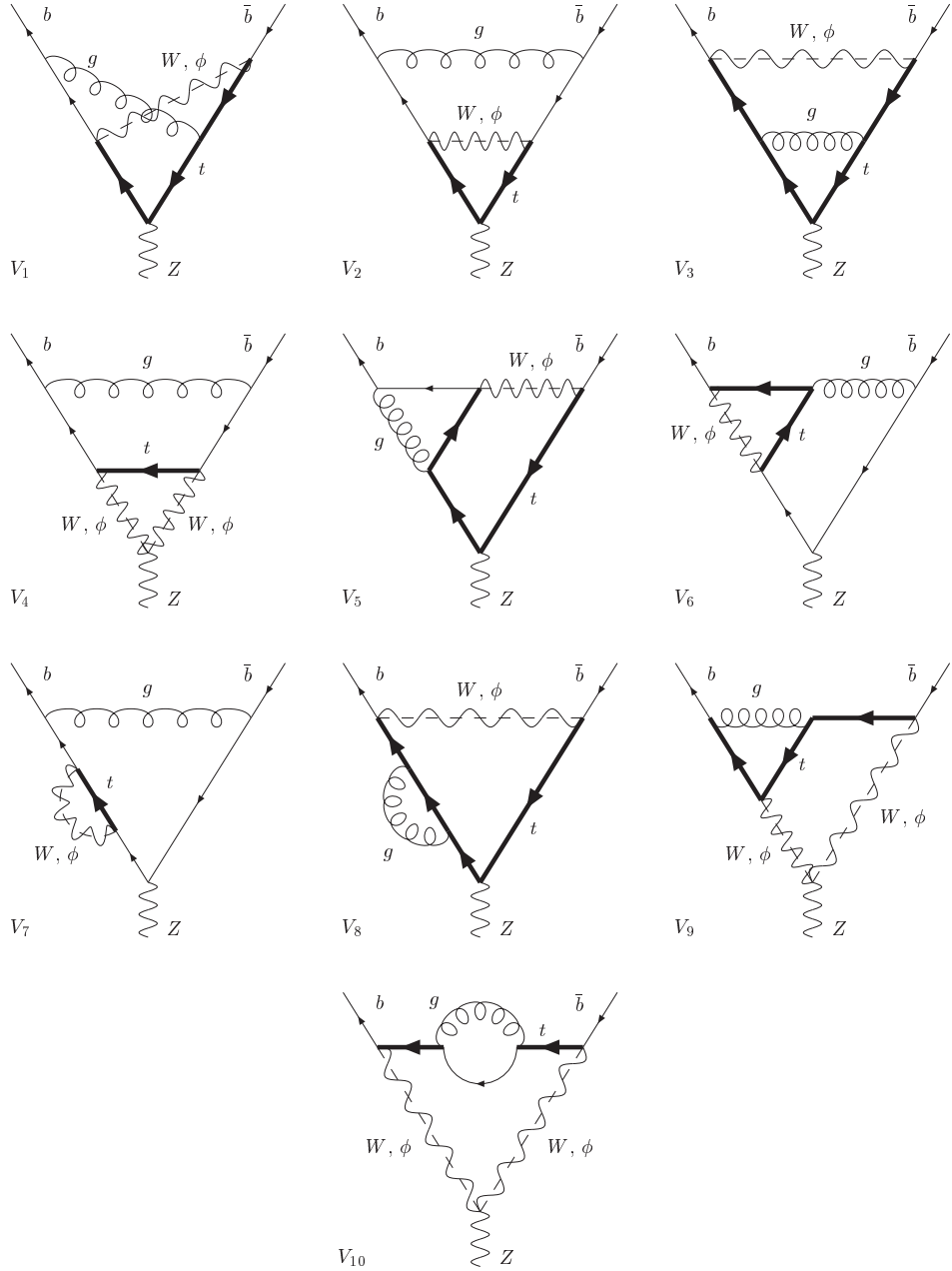}
\end{center}
\caption{Two-loop three-point diagrams contributing to 
            $Z \rightarrow b\bar b$ at ${\cal O}(\alpha_s g^2)$.}
\label{fig:diagrams}
\end{figure}

Here we would like to illustrate by means of 
a concrete example how the algorithm described above 
works in practice. Examples involving two-point functions
were given in ref. \cite{onshellexp,ztopdecay}. Here, we treat all two-loop 
three-point diagrams which contribute to the important physical process
$Z \rightarrow b \bar b$ at ${\cal O}(\alpha_s g^2)$. The diagrams
involved are shown in figure~\ref{fig:diagrams}. In the complete calculation
of this process, there are also self-energy type diagrams which contribute
to the $b$ wave function renormalization. These two-point function graphs
are simpler than the three-point graphs both analytically and numerically. 
Within our method, they were calculated in ref. \cite{onshellexp}. The 
process $Z \rightarrow b \bar b$ at ${\cal O}(\alpha_s g^2)$ at 
${\cal O}(\alpha_s g^2)$ was calculated by means of a mass expansion
in ref. \cite{topexpansion}.

\subsection{Isolating the IR singularities}

The two-loop methods we describe in this paper are intended primarily 
for the calculation of massive diagrams. However, sometimes massless 
particles are involved in a calculation and may lead to infrared 
singularities. We would like to stress that for purely massless calculations,
such as QCD radiative corrections, more efficient methods are available 
already in the literature, which make use of the absence of masses
\cite{qcd}. It is
often possible in the massless case to carry out the calculation completely
analytically. The discussion in this section applies to cases where the main
difficulty is related to the presence of several masses, while the infrared
structure is relatively simple. Typical examples are the purely electroweak and
the mixed electroweak-QCD radiative corrections, 
such as the $Z \rightarrow b \bar b$ at ${\cal O}(\alpha_s g^2)$
process discussed here.

The diagrams $V_2$, $V_4$, $V_6$, and $V_7$ are infrared divergent. The
most common treatment of IR divergencies uses dimensional regularization
to separate both the UV and the IR singularities. In our approach however,
this is not directly possible because in the expressions of eqs. 7 the 
$1/\epsilon$ expansion already separated the UV singularities, while the
possible IR singularities are still contained in the integral 
representations of the finite parts $h_i$, with the intention of calculating
these finite parts by numerical integration.

\begin{figure}
\hspace{1.cm}
    \epsfxsize = 10cm
\begin{center}
    \epsffile{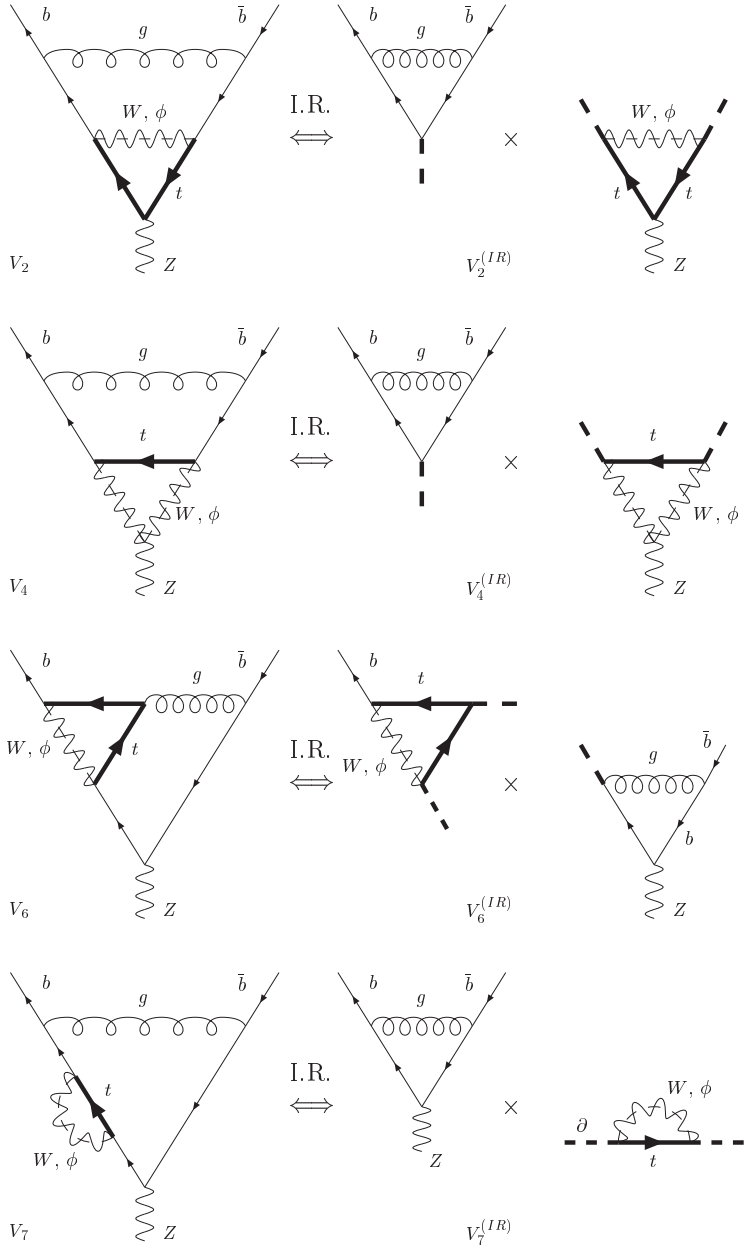}
\end{center}
\caption{{\em Extracting the infrared divergent pieces of the two-loop
              diagrams analytically. The infrared divergency of the two-loop
              diagram is the same as the infrared divergency of the 
              product of the two one-loop diagrams
              obtained by ``freezing'' the common line in the loop momenta 
              integration.}}
\label{fig:IR}
\end{figure}

Therefore it is useful to separate first the infrared part of the two-loop
diagrams in an analytically manageable form (one-loop diagrams in this 
case). The ``real two-loop'' calculation which is left after this separation
is then free of infrared singularities. 
The analytical separation of infrared divergencies is performed by 
noticing that the IR behaviour of the two-loop diagrams comes
essentially only from the loop integration over the gluon propagator.
Then, the IR behaviour is left unchanged
if the loop momentum on the propagator common to the two loop integrations
($r$) is being ``frozen'' to the loop momentum of the IR-finite loop.
This analytical separation of the IR behaviour is given in 
figure~\ref{fig:IR} for all IR divergent two-loop diagrams involved in
this process.

Once the IR separation is performed, the IR finite part 
($V_2-V_2^{(IR)}$ etc.) can be calculated by numerical integration.
For the numerical integration to be stable, one must make sure that
the two components, {\em e.g.} $V_2$ and $V_2^{(IR)}$, have the same
Feynman parameterization such that the singularities cancel already
before the Feynman parameters integration.
Then, the IR divergent part({\em e.g.} $V_2^{(IR)}$) 
can easily be treated analytically, in general by using 
dimensional regularization to regularize the IR singularities. 

For the
case of the process considered in this example, a gluon mass regulator
can also be used. This is possible because in this particular 
order in $\alpha_s$ the IR structure is the same as in the Abelian case,
and a gluon mass regulator can be used without upsetting the Slavnov-Taylor
identities.

It is still an open question if such an analytical separation of the 
infrared singularities can {\em always} be performed such that the numerical
integration can be carried out, in the case of other processes with potentially
more complicated IR structure. This question clearly requires 
further investigation.

\subsection{Numerical results}

We subtract the UV divergence of the diagrams of figure ~\ref{fig:diagrams} 
by performing minimal subtractions of overall-divergences and sub-divergences.
Further, we extract the IR divergences analytically, according to 
the discussion of the previous section. The tensor decomposition results in
a set of convergent integrals over the remaining Feynman parameters, which
can be carried out numerically. 

\begin{figure}
\hspace{1.cm}
\begin{center}
    \epsfxsize = 12.cm
    \epsffile{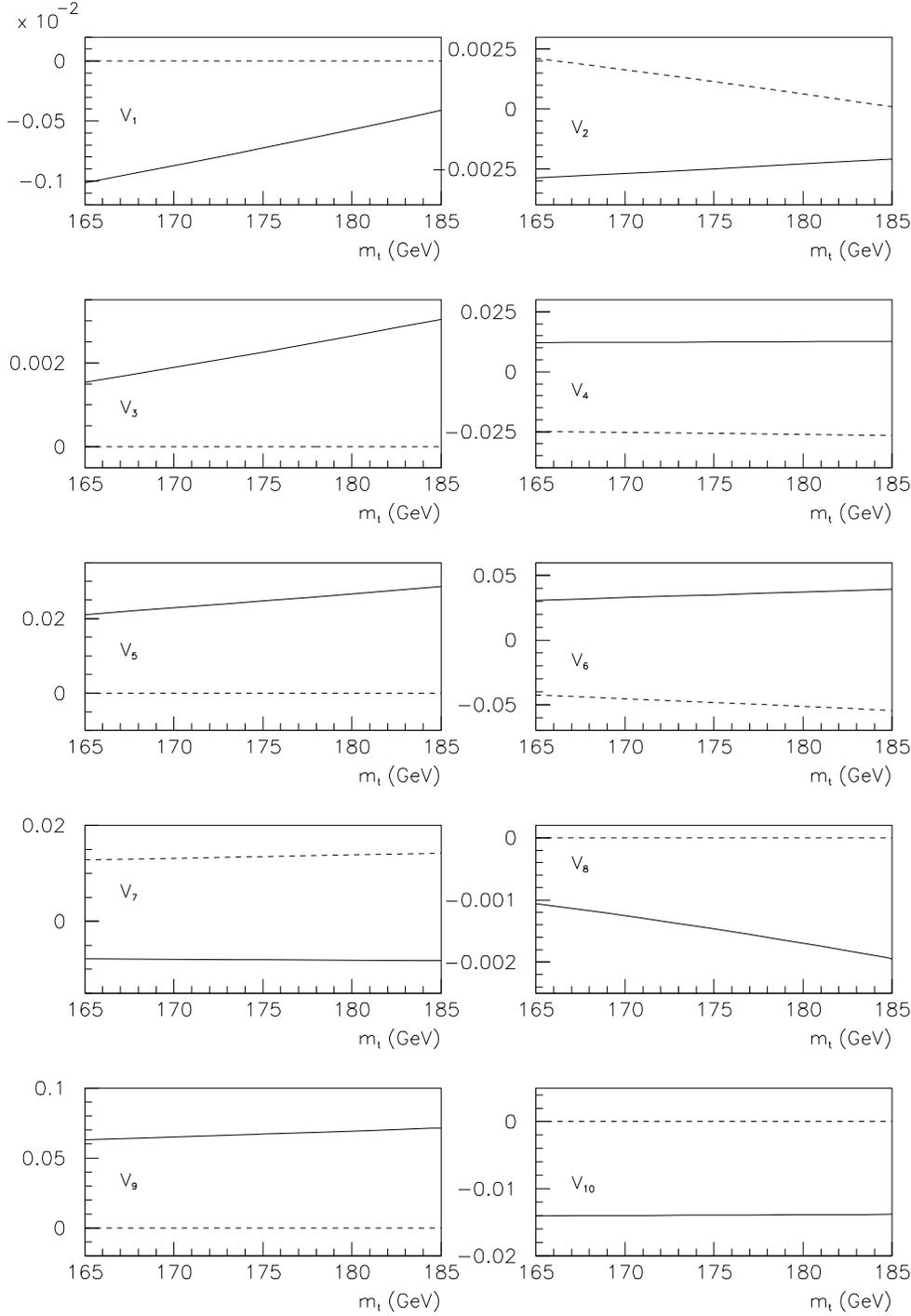}
\end{center}
\caption{{\em Numerical results for the two-loop three-point function
              diagrams shown in figure~\ref{fig:diagrams}. The UV
              divergences and sub-divergences are subtracted by 
              minimal subtraction. The IR divergences are subtracted
              according to figure~\ref{fig:IR}, so that the final result
              for diagrams $V_2$, $V_4$, $V_6$, and $V_7$ are the sum
              of the finite parts given in the plots plus the IR divergent
              one-loop contributions given in figure~\ref{fig:IR}.
              There is an overall factor of 
              $i \gamma_{\mu} (1-\gamma_5) \alpha_s (g^3/12 \cos \theta_W)$.
              The solid line is the real part, and the dashed line 
              is the imaginary part.}}
\label{fig:diagramplots}
\end{figure}

Because in the top mass range of physical interest the diagrams are not
close to a threshold, the top mass dependency in the relevant range
is relatively smooth, as it can be seen in figure~\ref{fig:diagramplots}.
We give numerical values for these diagrams in table 1 for three values of
the top mass --- intermediary values can be well approximated by interpolation.
In performing the numerical integration, we took $M_W=80.41$ GeV, 
$M_Z=91.187$ GeV, and the effective electroweak mixing angle
$\sin^2 \theta_W = .232$. For simplifying both 
the analytical and the numerical work, we neglected the $b$ mass, 
and in this limit all diagrams become
proportional to $\gamma_{\mu} (1-\gamma_5)$. It is perfectly 
possible to keep the exact $m_b$ dependency throughout the whole 
calculation at the price of dealing with longer expressions and more
computing time; 
however the $b$ mass effect on the two-loop $Z \rightarrow b \bar b$ decay 
rate is expected to be very small.

The numerical results are given in table 1. As for the numerical
accuracy and efficiency which can be attained, the results of table 1
are for a total of 26 individual Feynman diagrams 
($W$ and $\phi$ exchange counted separately),
each of them evaluated for 3 different values of 
the top mass. To carry out the numerical
integration for this total of 78 two-loop individual Feynman 
graph evaluations with an accuracy of $10^{-3}$,
a total of 100 hours computing time
on a 600 MHz Pentium machine was used. Higher numerical accuracy 
can be obtained by simply using more computing power.
The analytical tensor decomposition part, performed by a FORM
computer algebra program (we also used a Schoonship version with similar results),
takes about one hour.

\begin{table} 
\begin{tabular}{||l||c|c|c||}    \hline\hline
diagram           & $m_t=165$ GeV                       & $m_t=175$ GeV                       & $m_t=185$ GeV                       \\ \hline\hline
$V_1$             & $ -1.009            \cdot 10^{-3}$  & $-7.187             \cdot 10^{-4}$  & $ -4.057            \cdot 10^{-4}$  \\
$V_2-V_2^{(IR)}$  & $(-2.873 + i 2.122) \cdot 10^{-3}$  & $(-2.490 + i 1.147) \cdot 10^{-3}$  & $(-2.087 + i .09274)\cdot 10^{-3}$  \\
$V_3$             & $  1.545            \cdot 10^{-3}$  & $ 2.255             \cdot 10^{-3}$  & $  3.034            \cdot 10^{-3}$  \\
$V_4-V_4^{(IR)}$  & $( 1.215 - i 2.481) \cdot 10^{-2}$  & $( 1.242 - i 2.570) \cdot 10^{-2}$  & $( 1.266 - i 2.660) \cdot 10^{-2}$  \\
$V_5$             & $  2.107            \cdot 10^{-2}$  & $ 2.469             \cdot 10^{-2}$  & $  2.861            \cdot 10^{-2}$  \\
$V_6-V_6^{(IR)}$  & $( 3.089 - i 4.257) \cdot 10^{-2}$  & $( 3.500 - i 4.824) \cdot 10^{-2}$  & $( 3.950 - i 5.445) \cdot 10^{-2}$  \\
$V_7-V_7^{(IR)}$  & $(-.7778 + i 1.281) \cdot 10^{-2}$  & $(-.8001 + i 1.349) \cdot 10^{-2}$  & $(-.8232 + i 1.420) \cdot 10^{-2}$  \\
$V_8$             & $ -1.059            \cdot 10^{-3}$  & $-1.474             \cdot 10^{-3}$  & $ -1.942            \cdot 10^{-3}$  \\
$V_9$             & $  6.289            \cdot 10^{-2}$  & $ 6.703             \cdot 10^{-2}$  & $  7.143            \cdot 10^{-2}$  \\
$V_{10}$          & $ -1.402            \cdot 10^{-2}$  & $-1.389             \cdot 10^{-2}$  & $ -1.380            \cdot 10^{-2}$  \\ \hline\hline
\end{tabular} 
\caption{Numerical values for the two-loop diagrams shown in figure~\ref{fig:diagrams}. 
$V_1$--$V_{10}$ are the sums of the corresponding $W$ and $\phi$ 
exchange graphs.
An overall color and coupling constant factor of 
$i \gamma_{\mu} (1-\gamma_5) \alpha_s (g^3/12 \cos \theta_W)$ is understood.
The UV and IR divergences are removed  
as discussed in the text. The numerical integration accuracy is  
$10^{-3}$. The evaluation of a total of 78 Feynman graph evaluations with this precision
requires 100 hours computing time on a 600 MHz Pentium machine.}
\end{table}

\section{Conclusions}

We described an algorithm for the tensor reduction of massive two-loop
diagrams. It applies in principle to any massive two-loop graph, and it can
be automatized in the form of a computer algebra program. The tensor 
decomposition algorithm results in a set of ten special functions $h_i$ which
are defined in terms of one-dimensional integral representations.
We described the numerical methods which are used for carrying out the remaining
integrations. 

By applying the analytical reduction and numerical integration to an
important three-point example, $Z \rightarrow b \bar b$, 
we have shown that it can be used in realistic
calculations, where several internal mass and external momenta scales 
are involved. This approach works for any such combination of kinematic 
variables, apart maybe from possible infrared complications.

In the context of the $Z \rightarrow b \bar b$ example given in this paper,
we discussed the analytical separation of the infrared divergencies.
Within our two-loop methods, if a process involves infrared singularities,
these have to be dealt with in a special way because the numerical
nature of our methods. It is an
open question if this IR treatment can be generalized to other two-loop 
situations with potentially more complicated IR structure.

\vspace{1cm}

{\bf Aknowledgements}

The work of A.G. was supported by the US Department of Energy.
The work of Y.-P. Y. was supported partly by the US Department of Energy.

\section*{Appendix A}

In section 2.2 we introduced two types of scalar functions which are involved
in the tensor decomposition: ${\cal H}_i$ and 
$\tilde {\cal P}^{a b}_{\alpha_1 \alpha_2 \alpha_3}$. Then, we gave integral
representations only for ${\cal H}_i$. The necessary
$\tilde {\cal P}^{a b}_{2 \, 1 \, 1}$
scalar functions can be derived from the corresponding ${\cal H}_i$ by
partial fractioning:

\begin{eqnarray}
    \tilde{{\cal P}}^{0 \, 0}_{2 \, 1 \, 1} & = &
    {\cal H}_1
  \nonumber \\
    \tilde{{\cal P}}^{1 \, 0}_{2 \, 1 \, 1} & = &
    {\cal H}_2
  \nonumber \\
    \tilde{{\cal P}}^{0 \, 1}_{2 \, 1 \, 1} & = &
    {\cal H}_3
  \nonumber \\
    \tilde{{\cal P}}^{2 \, 0}_{2 \, 1 \, 1} & = &
    {\cal H}_4 + \frac{k^2}{n} \left[
  - (m_1^2 + k^2) {\cal H}_1
  - 2 {\cal H}_2
  + \tilde {\cal P}^{0 \, 0}_{1 \, 1 \, 1}
                              \right]
  \nonumber \\
    \tilde{{\cal P}}^{1 \, 1}_{2 \, 1 \, 1} & = &
    {\cal H}_5 + \frac{k^2}{2 n} \left\{
    (m_1^2 + m_2^2 - m_3^2 + k^2) {\cal H}_1
  + 2 {\cal H}_2
  - \tilde {\cal P}^{0 \, 0}_{1 \, 1 \, 1}   \right.
  \nonumber \\
   &   &   \left.
  \; \; \; \; \; \; \; \; \; \; 
  \; \; \; \; \; \; \; \; \; \; 
  \; \; \; \; \; \; \; \; \; \; 
  \; \; \; \; \; \; \; \; \; \; 
  \; \; \; \; \; \; \; \; \; \; 
  \; \; \; \; \; \; \; \; \; \; 
  + T_2(m_1^2) \left[ T_1(m_2^2) - T_1(m_3^2) \right]
                              \right\}
  \nonumber \\ 
    \tilde{{\cal P}}^{0 \, 2}_{2 \, 1 \, 1} & = &
    {\cal H}_6 + \frac{k^2}{n} \left[
  - m_2^2  {\cal H}_1
  + T_2(m_1^2) T_1(m_3^2) 
                              \right]
  \nonumber \\
    \tilde{{\cal P}}^{3 \, 0}_{2 \, 1 \, 1} & = &
    {\cal H}_7 + \frac{3 k^2}{n+2} \left[
  - (m_1^2 + k^2) {\cal H}_2
  + \tilde{{\cal P}}^{1 \, 0}_{1 \, 1 \, 1}
  - 2 \tilde{{\cal P}}^{2 \, 0}_{2 \, 1 \, 1}
                              \right]
  \nonumber \\
    \tilde{{\cal P}}^{2 \, 1}_{2 \, 1 \, 1} & = &
    {\cal H}_8 + \frac{3 k^2}{n+2} \left[
  - (m_1^2 + k^2) {\cal H}_3
  + {\cal P}^{0 \, 1}_{1 \, 1 \, 1}
  - 2 \tilde{{\cal P}}^{1 \, 1}_{2 \, 1 \, 1}
                              \right]
  \nonumber \\
    \tilde{{\cal P}}^{1 \, 2}_{2 \, 1 \, 1} & = &
    {\cal H}_9 + \frac{k^2}{n+2} \left[
    (m_1^2 + m_2^2 - m_3^2 + k^2) {\cal H}_3
  - m_2^2 {\cal H}_2
  + 2 \tilde{{\cal P}}^{1 \, 1}_{2 \, 1 \, 1}
  - \tilde {\cal P}^{0 \, 1}_{1 \, 1 \, 1}  \right.
  \nonumber \\
   &   &   \left.
  \; \; \; \; \; \; \; \; \; \;\; \; \; \; \; 
  \; \; \; \; \;\; \; \; \; \; \; \; \; \; \; 
  \; \; \; \; \; \; \; \; \; \;\; \; \; \; \; 
  \; \; \; \; \;\; \; \; \; \; \; \; \; \; \; 
  \; \; \; \; \;\; \; \; \; \; \; \; \; \; \; 
  - 2 k^2 T_2(m_1^2) T_1(m_3^2) 
                              \right]
  \nonumber \\
    \tilde{{\cal P}}^{0 \, 3}_{2 \, 1 \, 1} & = &
    {\cal H}_{10} + \frac{3 k^2}{n+2} \left[
  - m_2^2 {\cal H}_3
  + k^2 T_2(m_1^2) T_1(m_3^2)     
                              \right]
  \; \; \; \; ,
\end{eqnarray}
where $T_1$ and $T_2$ are the Euclidian one--loop tadpole integrals:

\begin{eqnarray}
   T_1(m^2) & = & \int d^{n}p\, \frac{1}{p^2+m^2}  
    \; = \; 
   - \pi^2 
     \left( \pi m^2 \right)^{\frac{\epsilon}{2}}
     \Gamma\left( - \frac{\epsilon}{2} \right) \,
     \frac{2 m^2}{2+\epsilon} 
  \nonumber \\
   T_2(m^2) & = & \int d^{n}p\, \frac{1}{(p^2+m^2)^2}  
    \; = \; 
     \pi^2 
     \left( \pi m^2 \right)^{\frac{\epsilon}{2}}
     \Gamma\left( - \frac{\epsilon}{2} \right)
    \; \; .
\end{eqnarray}

In section 2.4, in order to simplify the discussion, we introduced 
the functions ${\cal P}^{a b}_{\alpha_1 \alpha_2 \alpha_3}$
which are related to $\tilde {\cal P}^{a b}_{\alpha_1 \alpha_2 \alpha_3}$
by a simple loop momentum shift:

\begin{eqnarray}
   {\cal P}^{0 \, 0}_{2 \, 1 \, 1} & = &
   \tilde{{\cal P}}^{0 \, 0}_{2 \, 1 \, 1}
  \nonumber \\
   {\cal P}^{1 \, 0}_{2 \, 1 \, 1} & = &
 -     \tilde{{\cal P}}^{1 \, 0}_{2 \, 1 \, 1}
 - k^2 \tilde{{\cal P}}^{0 \, 0}_{2 \, 1 \, 1}
  \nonumber \\
   {\cal P}^{0 \, 1}_{2 \, 1 \, 1} & = &
   \tilde{{\cal P}}^{0 \, 1}_{2 \, 1 \, 1}
  \nonumber \\
   {\cal P}^{2 \, 0}_{2 \, 1 \, 1} & = &
           \tilde{{\cal P}}^{2 \, 0}_{2 \, 1 \, 1}
 + 2 k^2   \tilde{{\cal P}}^{1 \, 0}_{2 \, 1 \, 1}
 + (k^2)^2 \tilde{{\cal P}}^{0 \, 0}_{2 \, 1 \, 1}
  \nonumber \\
   {\cal P}^{1 \, 1}_{2 \, 1 \, 1} & = &
 -     \tilde{{\cal P}}^{1 \, 1}_{2 \, 1 \, 1}
 - k^2 \tilde{{\cal P}}^{0 \, 1}_{2 \, 1 \, 1}
  \nonumber \\
   {\cal P}^{0 \, 2}_{2 \, 1 \, 1} & = &
   \tilde{{\cal P}}^{0 \, 2}_{2 \, 1 \, 1}
  \nonumber \\
   {\cal P}^{3 \, 0}_{2 \, 1 \, 1} & = &
 -           \tilde{{\cal P}}^{3 \, 0}_{2 \, 1 \, 1}
 - 3  k^2    \tilde{{\cal P}}^{2 \, 0}_{2 \, 1 \, 1}
 - 3 (k^2)^2 \tilde{{\cal P}}^{1 \, 0}_{2 \, 1 \, 1}
 -   (k^2)^3 \tilde{{\cal P}}^{0 \, 0}_{2 \, 1 \, 1}
  \nonumber \\
   {\cal P}^{2 \, 1}_{2 \, 1 \, 1} & = &
            \tilde{{\cal P}}^{2 \, 1}_{2 \, 1 \, 1}
 + 2 k^2    \tilde{{\cal P}}^{1 \, 1}_{2 \, 1 \, 1}
 +  (k^2)^2 \tilde{{\cal P}}^{0 \, 1}_{2 \, 1 \, 1}
  \nonumber \\
   {\cal P}^{1 \, 2}_{2 \, 1 \, 1} & = &
 -     \tilde{{\cal P}}^{1 \, 2}_{2 \, 1 \, 1}
 - k^2 \tilde{{\cal P}}^{0 \, 2}_{2 \, 1 \, 1}
  \nonumber \\
   {\cal P}^{0 \, 3}_{2 \, 1 \, 1} & = &
   \tilde{{\cal P}}^{0 \, 3}_{2 \, 1 \, 1}
    \; \; .
\end{eqnarray}

The functions with $\alpha_1=\alpha_2=\alpha_3=1$ which appear in
the relations 18 can be calculated by partial $p$ (eq. 15).
For simplifying the notation, we omitted in the above formulae
the mass and momentum arguments of the functions
${\cal H}_i(m_1,m_2,m_3;k^2)$,
${\cal P}^{a \, b}_{\alpha_1\, \alpha_2 \, \alpha_3}
(m_1,m_2,m_3;k^2)$, and
$\tilde{{\cal P}}^{a \, b}_{\alpha_1\, \alpha_2 \, \alpha_3}
(m_1,m_2,m_3;k^2)$, and understand that these arguments appear in 
this same order in all relations.

\section*{Appendix B}

The integral representations of the ten special functions functions $h_i$,
given in eqs. 8, are built from the following functions:

\begin{eqnarray}
  \tilde{g} (m_1,m_2,m_3;k^2;x) & = &
     Sp(\frac{1}{1-y_1}) 
   + Sp(\frac{1}{1-y_2}) 
   + y_1 \log{\frac{y_1}{y_1-1}} 
   + y_2 \log{\frac{y_2}{y_2-1}} 
  \nonumber \\
  \tilde{f_1}(m_1,m_2,m_3;k^2;x) & = &
   \frac{1}{2}
   \left[
   - \frac{1-\mu^2}{\kappa^2}
   + y_1^2 \log{\frac{y_1}{y_1-1}} 
   + y_2^2 \log{\frac{y_2}{y_2-1}} 
   \right]
  \nonumber \\
  \tilde{f_2}(m_1,m_2,m_3;k^2;x) & = &
   \frac{1}{3}
   \left[
   - \frac{2}{\kappa^2} 
   - \frac{1-\mu^2}{2 \kappa^2}
   - \left( \frac{1-\mu^2}{\kappa^2} \right)^2
   \right.
  \nonumber \\
 & &
   \; \; \; \; \; \; \; \; \; \; \; \;
   \; \; \; \; \; \; \; \; \; \; \; \;
   \; \; \; \; \; \; \; \; \; \; \; \;
   \left.
   + y_1^3 \log{\frac{y_1}{y_1-1}} 
   + y_2^3 \log{\frac{y_2}{y_2-1}} 
   \right]
  \nonumber \\
  \tilde{f_3}(m_1,m_2,m_3;k^2;x) & = &
   \frac{1}{4}
   \left[
   - \frac{4}{\kappa^2} 
   - \left( \frac{1}{3} + \frac{3}{\kappa^2}  \right) 
     \left( \frac{1-\mu^2}{\kappa^2} \right)
   - \frac{1}{2} \left( \frac{1-\mu^2}{\kappa^2} \right)^2
   - \left( \frac{1-\mu^2}{\kappa^2} \right)^3
   \right.
  \nonumber \\
 & &
   \left.
   \; \; \; \; \; \; \; \; \; \; \; \;
   \; \; \; \; \; \; \; \; \; \; \; \;
   \; \; \; \; \; \; \; \; \; \; \; \;
   + y_1^4 \log{\frac{y_1}{y_1-1}} 
   + y_2^4 \log{\frac{y_2}{y_2-1}} 
   \right]
   \; \; ,
\end{eqnarray}
where we use the following notations:

\begin{eqnarray}
y_{1,2} & = & \frac{1 + \kappa^{2} - \mu^{2}
                    \pm \sqrt{\Delta}}{2 \kappa^{2}}  \nonumber \\
\Delta  & = & (1 + \kappa^{2} - \mu^{2})^{2} 
          + 4 \kappa^{2} \mu^{2} - 4 i \kappa^{2} \eta 
      \; \; ,
\end{eqnarray}
and

\begin{eqnarray}
   \mu^{2}  & = &  \frac{a x + b (1-x)}{x (1-x)}   \nonumber \\
         a  & = &  \frac{m_{2}^{2}}{m_{1}^{2}} \, , \; \; \; \;
         b \; = \; \frac{m_{3}^{2}}{m_{1}^{2}} \, , \; \; \; \;
\kappa^{2} \; = \; \frac{    k^{2}}{m_{1}^{2}} 
      \; \; .
\end{eqnarray}

In the above expressions, one special case must be treated separately,
namely $k^2 = 0$. The $h_i$ functions have a smooth limit for 
$k^2 \rightarrow 0$. For the purpose of numerical evaluation, it is useful to
use a Taylor expansion of the functions $\tilde g$, $\tilde f_1$, 
$\tilde f_2$, and $\tilde f_3$ around
$k^2=0$ for extremely small values of $k^2$, where a direct evaluation
by means of the exact expressions given above would be affected by large 
cancellations.  It can be checked that this limit is regular and
our approach reduces to the functions introduced by van der Bij 
and Veltman in ref. \cite{vdbij}.



\end{document}